# Glassforming liquid crystalline equimolar mixtures of MHPOBC and fluorinated compounds – structural, optical, and dielectric properties

Aleksandra Deptuch[1,*], Anna Paliga[2], Anna Drzewicz[1], Michał Czerwiński[3], Ewa Juszyńska-Gałązka[1,4]

[1] Institute of Nuclear Physics Polish Academy of Sciences, Radzikowskiego 152, PL-31342 Kraków, Poland

[2] Faculty of Physics and Applied Computer Science, AGH University of Kraków, Reymonta 19, PL-30059 Kraków, Poland

[3] Institute of Chemistry, Military University of Technology, Kaliskiego 2, PL-00908 Warsaw, Poland

[4] Research Center for Thermal and Entropic Science, Graduate School of Science, Osaka University, 560-0043 Osaka, Japan

*corresponding author, aleksandra.deptuch@ifj.edu.pl

**Abstract**

Three liquid crystalline equimolar mixtures are formulated, each containing the MHPOBC compound and the homolog from the 3FmHPhF6 series with m = 4, 5, and 6. The properties of the mixtures are investigated by differential scanning calorimetry, polarizing optical microscopy, X-ray diffraction, electro-optic measurements, and UV-Vis-NIR and dielectric spectroscopies. In all mixtures, the smectic A*, smectic C*, and smectic $C_A$* phases are observed. Each mixture can be supercooled and forms the glass of smectic $C_A$* at 225-238 K. The glass transition temperature decreases with the increasing $C_mH_{2m}$ chain length in the 3FmHPhF6 molecule. The tilt angle in the smectic $C_A$* phase reaches 36.5-37.5°. The selective reflection of light in the visible and near-infrared spectral ranges is observed, with a strong dependence of the reflected light's wavenumber on temperature.

## 1. Introduction

The first reports of the ferro- and antiferroelectric properties in the liquid crystals were published for the DOBAMBC compound in 1975 [1] and for the MHPOBC compound in 1989 [2], respectively. Such properties can be exhibited, under certain conditions, by the tilted smectic phases, where the molecules are ordered into layers and tilted on average by an angle $\theta > 0$ from the smectic layer normal. The molecules are required to be chiral in order to decrease the symmetry and enable a non-zero spontaneous polarization of the smectic layer [1]. The helical order of the tilt direction, present for the chiral molecules, has to be removed in the surface-stabilized bookshelf geometry to provide the non-zero spontaneous polarization of the bulk sample, consisting of numerous smectic layers. Then, the bistable and tristable switching by the electric field, indicating ferro- and antiferroelectricity, respectively, can be observed in the smectic C* (SmC*) and smectic $C_A$* (Sm$C_A$*) phases with the synclinic and anticlinic order of the tilt direction in the neighbor layers [1-4]. The * notation indicates the chirality of molecules. The exceptions from these conditions are the achiral bent-core mesogens or bent-shaped dimers, where the ferro- and antiferroelectricity can be obtained not only in SmC, but also in the SmA phase, where the average tilt angle $\Theta = 0$ [5,6]. The ferro- and antiferroelectric properties are reported even for some achiral rod-like molecules [7], however, this study involves 'classical' liquid crystalline ferro/antiferroelectrics, as described in [1,2]. The discoveries of the bistable and tristable switching in chiral tilted smectics opened the possibility of the liquid crystal displays based on the smectic phases [8-12]. The helical pitch in the SmC* and SmC$_A$* phases is typically of 100-1000 nm order of magnitude, which enables the selective reflection of visible light by the homeotropically aligned samples. The helical pitch is usually sensitive to temperature, giving such liquid crystals the thermochromic property [13-16]. Some mesogenic compounds form the glass of various liquid crystalline phases, presumed for optical or electronic applications [17-19]. Tuning the properties of liquid crystals: the stability range of the chiral tilted smectic phases, electro-optic response, and helical pitch, to the values desirable in a given device, is performed usually by formulation of mixtures [20-23].

This study is dedicated to the equimolar liquid crystalline mixtures of (*S*)-4-[(1-methylheptyloxy)carbonyl]phenyl 4′-octyloxy-4-biphenylcarboxylate, known as MHPOBC, with the homologs from the series of (*S*)-4'-(1-methylheptyloxycarbonyl)biphenyl-4-yl 4-[m-(2,2,3,3,4,4,4-heptafluorobutoxy)alkyl-1-oxy]-2-fluorobenzoate, denoted as 3FmHPhF6, with m = 4, 5, 6 (Figure 1). MHPOBC forms the SmA* and SmC$_A$* phases. Between them, in a narrow temperature range, there can exist the SmC* phase and various sub-phases: SmC$_\alpha$*, SmC$_{FI1}$*, and SmC$_{FI2}$*, which presence or absence depends strongly on the optical purity of the sample [24]. The supercooled MHPOBC undergoes the transition to the SmX$_A$* phase, which possesses a hexatic order, i.e., all domains with the short-range hexagonal order show the same orientation within the smectic layer [25,26]. The recent dielectric spectra of the MHPOBC sample used in this study suggest the sequence SmA* → SmC$_\alpha$* → SmC* → SmC$_{FI2}$* → SmC$_A$* → SmX$_A$* [27]. The 3FmHPhF6 compounds form the SmC*, SmC$_A$* phases for m = 4 and SmA*, SmC*, and SmC$_A$* phases for m = 5, 6 [28-30]. The homolog with m = 4 crystallizes on cooling and the critical cooling rate necessary to obtain the glassy SmC$_A$* phase is estimated to 60-80 K/min [31]. Meanwhile, the glass transition of the SmC$_A$* phase is observed for m = 5, 6 at much slower cooling [32].

The aim of this paper is testing the properties of three MHPOBC/3FmHPhF6 equimolar mixtures, in particular, their tendency to form the glassy smectic phase, switching in the electric field and selective reflection, which are crucial for their potential applications. The phase transitions of mixtures are investigated by differential scanning calorimetry (DSC) and polarizing optical microscopy (POM). The structural parameters of the smectic phases are determined by X-ray diffraction (smectic layer spacing, average intermolecular distance, correlation length), electro-optic measurements (tilt angle, spontaneous polarization) and UV-Vis-NIR spectrometry (helical pitch). The relaxation processes, including the α-process related to the glass transition, is investigated by broadband dielectric spectroscopy (BDS). The influence of the $C_mH_{2m}$ chain length in the 3FmHPhF6 homolog on the mixture's properties is determined.

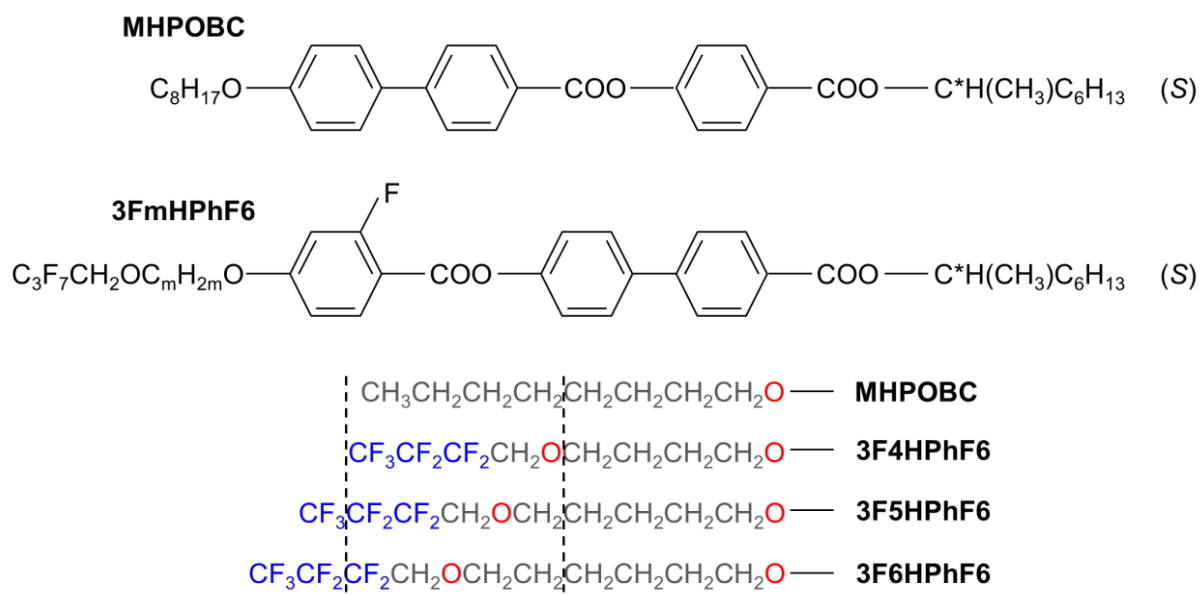

Figure 1. Molecular structures of the MHPOBC compound and 3FmHPhF6 homologs, and the schematic comparison of their achiral chain lengths.

## 2. Experimental details

The MHPOBC [24] and 3FmHPhF6 (m = 4, 5, 6) [28-30] compounds were synthesized in the Institute of Chemistry of the Military University of Technology in Warsaw. All compounds are S-enantiomers. Three mixtures were prepared:

- MIX4HF6 – MHPOBC : 3F4HPhF6, molar ratio 0.4980(4) : 0.5020(5)
- MIX5HF6 – MHPOBC : 3F5HPhF6, molar ratio 0.4935(2) : 0.5065(2)
- MIX6HF6 – MHPOBC : 3F6HPhF6, molar ratio 0.5008(2) : 0.4992(2)

The components of each mixture were dissolved in acetone and mixed in a solution. After the evaporation of acetone, the precipitate was heated to the isotropic liquid phase (~433 K) and cooled back to the room temperature.

The DSC measurements were performed with DSC 2500 TA Instruments calorimeter for the samples in aluminum pans. The samples weighted 4.20, 6.68 and 9.36 mg for MIXmHF6 with m = 4, 5, 6, respectively. The thermograms were collected in the 173-433 K range, for the cooling/heating rates of 2-40 K/min. The DSC data were analyzed with TRIOS and OriginPro.

The POM textures were collected with 10× magnitude under the Leica DM2700 P polarizing microscope equipped with the Linkam temperature stage. The samples were placed between glass slides without aligning layers. The POM observations were done in the 273-433 K range for the 5 K/min rate and in the 193-433 K for the 30 K/min rate. The average luminance of each texture was calculated in TOApy [33].

The XRD patterns were collected with the X'Pert PRO PANalytical diffractometer (CuKα radiation) in the Bragg-Brentano geometry for the flat samples in the 13 mm × 10 mm × 0.2 mm holder. The measurements were done on cooling in the 298-433 K range. The NIST Standard Reference Material 675 [34], supplied by Merck, was used for 2θ calibration. The XRD data were analyzed with FullProf [35] and OriginPro.

The electro-optic measurements were performed for samples in a surface-stabilized bookshelf geometry, in the alternating electric field provided by the R&S HMF2550 function generator and FLC F20AD amplifier, and temperature controlled by the Linkam THMS 600 stage under the Biolar PI – PZO Poland polarizing microscope. The tilt angle was determined by application of the rectangular field (30 Hz frequency, 30-50 V amplitude) and observation of the Clark-Lagerwall effect [3] with the Thorlabs PDA100A photodetector. The spontaneous polarization was determined by the reversal current method [36] by application of the triangular field (50 Hz, 30-50 V) and collection of the sample's response with the R&S HM0724 oscilloscope.

The UV-Vis-NIR spectra in the 360-3000 nm range were recorded with Shimadzu UV-Vis-NIR spectrometer with MLWU7 temperature controller for the homeotropically ordered samples. The helical pitch was determined as $p_h = \lambda_{min}/n_{av}$ [37,38], where $\lambda_{min}$ is the wavelength corresponding to the minimum in the transmitted light intensity and $n_{av}$ = 1.5 is the assumed refraction index [39].

The BDS spectra were recorded with Novocontrol impedance spectrometer for the samples with a thickness of 80 μm between two gold electrodes. The polytetrafluoroethylene spacers were applied to avoid the shortcut. The measurements were carried out for frequencies between 0.1 Hz and 10 MHz in the 173-433 K range. Two temperature programs were conducted: (1) measurement on gradual cooling and subsequent heating and (2) direct cooling at 10 K/min from 433 K to 173 K and measurement upon gradual heating. The spectra were analyzed by fitting a complex function in OriginPro.

## 3. Results and discussion
### 3.1. Phase sequence and glass transition

The DSC thermograms collected for the cooling/heating rates between 2 and 40 K/min are presented in Figure 2 and the corresponding phase transition temperatures are shown in Figure 3. The onset temperatures [40] of anomalies in the DSC thermograms are taken as the phase transition temperatures, except the glass transition, where the middle of the step in the heat capacity was taken. The DSC results indicate the same sequence of the smectic phases in the MIXmHF6 (m = 4, 5, 6) mixtures: SmA*, SmC*, and SmC$_A$* both on cooling and upon heating (enantiotropic phases). The monotropic hexatic SmX$_A$* phase is not observed. The temperature range of SmA* broadens slightly with the increasing C$_m$H$_{2m}$ chain length in the 3FmHPhF6 component. The temperature range of SmC* is broader for MIX4HF6 and MIX6HF6 than for MIX5HF6 (Table 1), which reflects the lower stability of SmC* in odd 3FmHPhF6 homologs [28-30].

MIX4HF6 and MIX6HF6 show partial crystallization for the cooling rates 2-20 K/min and 2-5 K/min, respectively, followed by the glass transition of the remaining $SmC_A^*$ fraction. At higher cooling rates, only the vitrification of $SmC_A^*$ is visible in the DSC thermograms. MIX5HF6 has the best glassforming ability because the melt crystallization is not observed even for the lowest 2 K/min rate and the $SmC_A^*$ glass is formed in the whole volume of the sample. The glass transition temperature $T_g$ decreases with the m parameter, both in the cooling and heating runs (Table 1). The cold crystallization is observed during heating for all mixtures, although for MIX5HF6, it does not occur for the 30-40 K/min heating rates. The number of the crystal phases visible in the DSC thermograms is four for MIX4HF6 and MIX6HF6, and three for MIX5HF6; the most stable crystal phases have the onset melting temperature equal to 329.5 K, 320 K, and 322 K for MIX4HF6, MIX5HF6, and MIX6HF6, respectively.

The POM textures and the luminance vs. temperature plots are gathered in Figures S1-S12 in the Supplementary Materials. The transitions between the smectic phases are easily recognizable in POM observations, while the glass transition of $SmC_A^*$ does not cause any significant changes in the textures. The $T_g$ values from the DSC results are indicated in the luminance vs. temperature plots; one can see that no anomaly in the luminance is visible around $T_g$. The results from Table 1 show that the transition temperatures obtained by POM are higher than determined by DSC; the smaller thickness and overall volume of the POM sample may contribute to this shift.

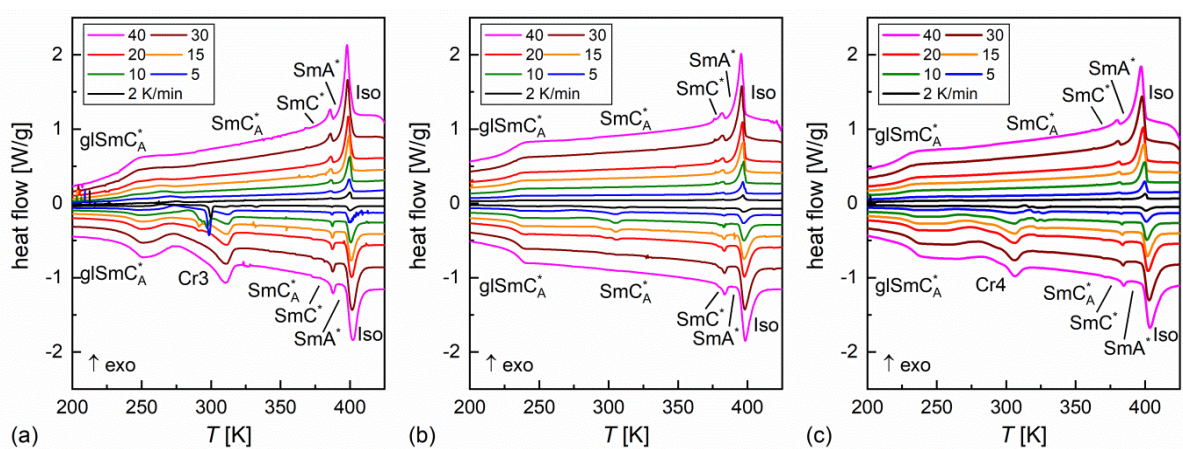

Figure 2. DSC thermograms of MIX4HF6 (a), MIX5HF6 (b), and MIX6HF6 (c). The upper and bottom lines are results from cooling and subsequent heating at the same rate, respectively.

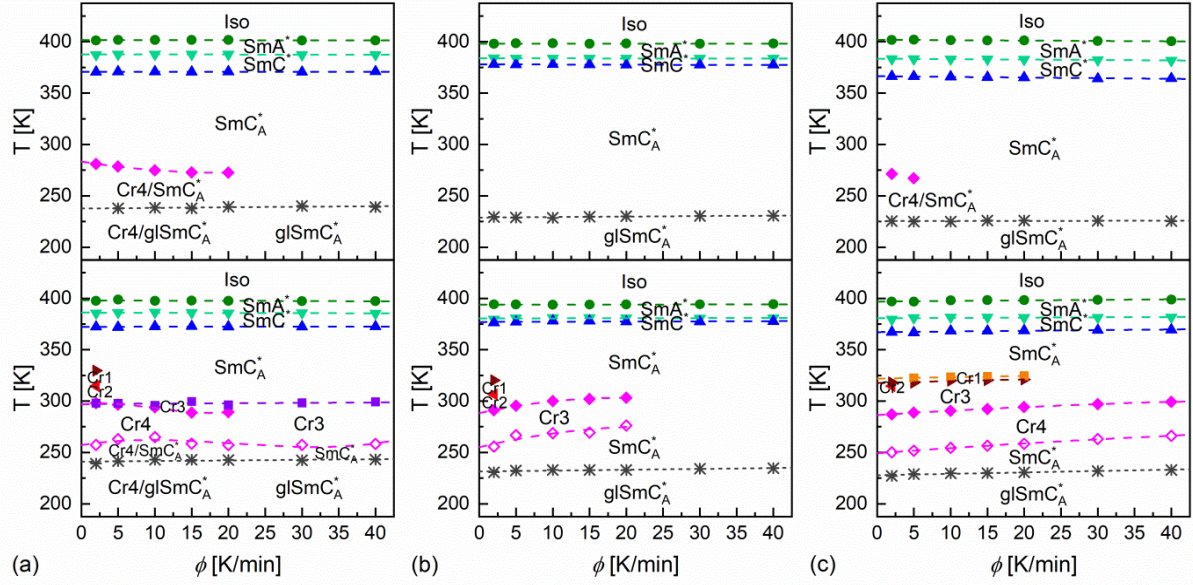

Figure 3. Phase transition temperatures vs the cooling/heating rates (upper/bottom panels) determined by DSC for MIX4HF6 (a), MIX5HF6 (b), and MIX6HF6 (c).

Table 1. Transitions between the smectic phases and glass transition of MIXmHF6. First row: transition temperatures in K determined by POM at 5 K/min. **Second row**: transition temperatures in K determined by DSC, extrapolation to 0 K/min. *Third row*: heat capacity change $\Delta C$ in kJ/(mol·K) at the glass transition (second column) and enthalpy changes $\Delta H$ in kJ/mol (columns 3-5) determined by DSC.

| transition | glSmC$_A$*/SmC$_A$* | SmC$_A$*/SmC* | SmC*/SmA* | SmA*/Iso |
|---|---|---|---|---|
| | cooling | | | |
| | - | 371 | 389 | 402 |
| MIX4HF6 | **237.7** | **370.8** | **387.7** | **401.7** |
| | *0.17* | *>0.1* | *0.6* | *5.4* |
| | - | 380 | 386 | 401 |
| MIX5HF6 | **228.9** | **378.0** | **383.8** | **398.2** |
| | *0.17* | *0.1* | *0.5* | *5.2* |
| | - | 368 | 386 | 404 |
| MIX6HF6 | **225.4** | **366.4** | **383.4** | **401.9** |
| | *0.17* | *>0.1* | *0.4* | *5.6* |
| | heating | | | |
| | - | 378 | 390 | 403 |
| MIX4HF6 | **240.9** | **372.3** | **386.0** | **398.1** |
| | *0.17* | *0.1* | *0.6* | *5.5* |
| | - | 381 | 386 | 401 |
| MIX5HF6 | **231.4** | **377.0** | **380.4** | **393.3** |
| | *0.17* | *0.1* | *0.5* | *5.5* |
| | - | 371 | 385 | 403 |
| MIX6HF6 | **227.8** | **367.1** | **380.6** | **397.0** |
| | *0.16* | *>0.1* | *0.4* | *5.9* |

### 3.2. Structure of smectic phases

The smectic A and C phases exhibit the quasi-long-range positional order: the molecules are arranged into layers [4,41]. It shows in the XRD patterns as a sharp peak or peaks at low angles (Figure 4a). The peak positions are related to the smectic layer spacing by the Bragg equation: $n\lambda = 2d \sin\theta$, where $n = 1, 2, 3...$ is the peak's order, $\lambda$ is the X-ray wavelength, $d$ is the layer spacing, and $\theta$ is the peak's position [42]. First three diffraction peaks ($n = 1, 2, 3$) recorded at 298 K were applied to remove any systematic shift in the peak positions by the method shown earlier in [31] (the samples were introduced to the holder in the crystal state and the slight volume change after melting could have led to the systematic shift, despite the earlier calibration with the crystalline SRM 675). The layer spacing increases with the increasing $C_mH_{2m}$ chain length in the 3FmHPhF6 component (Figure 4b). The layer shrinkage for m = 4, 5, 6 is equal to 2(1)%, 1.7(5)%, 1.4(8)%, respectively, at the SmA*/SmC* transition and 10(1)%, 8.8(4)%, 8.1(7)% in the whole investigated temperature range.

The isotropic liquid is characterized by the short-range positional order, which shows in the XRD patterns as a diffuse maximum at ~18° (Figure 4a). In the smectic A and C phases, the positional order within layers remains also short-range, although the maximum becomes narrower and shifts to higher angles due to increasing correlation length and decreasing average distance between molecules, respectively. Both parameters can be obtained by fitting the Lorentz peak function to the diffuse maximum in the scattering vector space [43]:

$$I(q) = \frac{A}{1+\xi^2(q-q_0)^2} + Bq + C, \qquad (1)$$

where: $A$ is the peak height, $B$ and $C$ describe the linear background, $q = 4\pi/\sin\theta$ is the scattering vector, $q_0$ is the peak position, and $\xi$ is the correlation length. The average intermolecular distance is obtained as $w = 2\pi/q_0$. It equals 4.5-5.0 Å in MIXmHF6 and decreases on cooling, except the SmA* phase of MIX5HF6, where it increases to almost 5.1 Å (Figure 4c). The correlation length equals 3.5-3.9 Å in the isotropic liquid phase and increases to 4.1-4.2 Å in SmA*. Deep in the SmC$_A$* phase, in 298-300 K, the correlation length reaches 5.5-5.8 Å in MIXmHF6 with m = 4, 5 and 4.9 Å in MIX6HF6. The $\xi/w$ ratio equals 0.7-1.2, thus, there are only next-neighbor correlations.

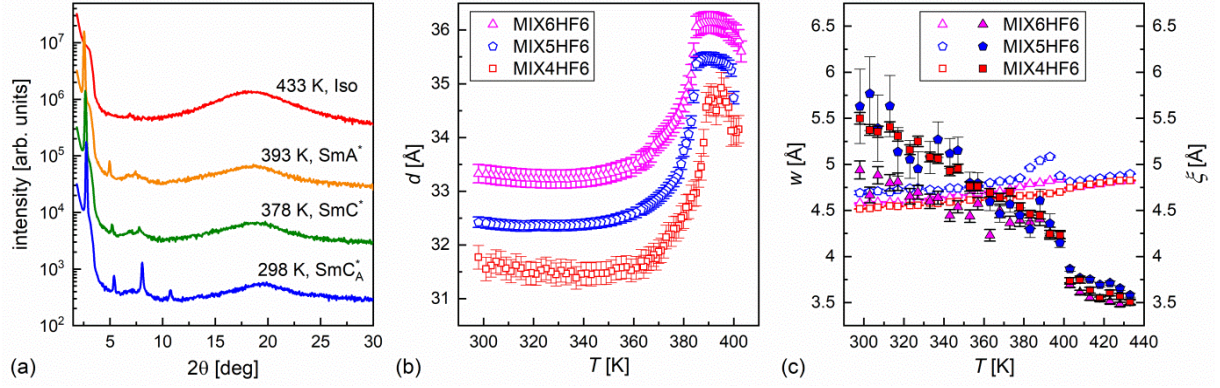

Figure 4. Selected XRD patterns of MIX6HF6 (a), smectic layer spacing for MIXmHF6 (b), and average intermolecular distance (open symbols) and correlation length (solid symbols) corresponding to the positional short-range order in the liquid and smectic phases of MIXmHF6 (c). The high intensity below 5° and a small peak at 7° in (a) are background contributions.

The tilt angle $\Theta$ has a maximal value of 36.5-37.5° for all mixtures (Figure 5a). The spontaneous polarization $P_s$ decreases slightly with the increasing $C_mH_{2m}$ chain length, with the maximal values 294, 283, 267 nC/cm$^2$ for m = 4, 5, 6, respectively (Figure 5b). The $P_s(T)$ dependence follows the power formula $P_0(T - T_c)^\gamma$ [44] in the whole temperature range, except two points at high temperatures for MIX5HF6. In contrary, the $\Theta(T)$ dependence follows the $\Theta_0(T - T_c')^{\gamma'}$ formula only close to the SmC*/SmA* transition, because it saturates on approaching the room temperature. The lack of proportionality between $\Theta$ and $P_s$ is a consequence of the high values of $\Theta$. In this case, the description of the coupling between these values requires using also the third power of $\Theta$ [44]. The parameter $\gamma$ for the spontaneous polarization equals 0.32, 0.39, 0.38 for m = 4, 5, 6, similarly as for high-tilted compounds investigated in [44]. Meanwhile, the corresponding $\gamma'$ parameter estimated for the tilt angle equals 0.12, 0.24, 0.31 for m = 4, 5, 6, respectively.

The relationship between the smectic layer spacing and tilt angle is described as [45]:

$$d = L\cos(\Theta - \delta\Theta), \qquad (2)$$

where: $L$ is the molecular length and $\delta\Theta$ is deviation from the linear shape, equal to zero for a rod-like molecule. The proposed formula is based on the assumption that the tilt angle measured by the electro-optic method corresponds to the tilt of the molecular core, which differs from the overall molecular tilt if the molecule has a non-linear shape [46-48]. The fitting of Equation (2) was performed separately for two temperature ranges (Figure 5c, Table 2). For MIX4HF6 and MIX6HF6, these overlap with the temperature ranges of the SmC* and SmC$_A$* phases, while for MIX5HF6, the border between fitted ranges is a few degrees below the SmC*/SmC$_A$* transition. The layer spacing increases slightly on cooling deep in the SmC$_A$* phase and these points were excluded from fitting, same as the points in the vicinity of the SmA*/SmC* transition. The $\Theta$ values were interpolated to match the temperature points where the layer spacing was determined. The fitting results show the decrease in $L$ and increase in $\delta\Theta$ in the SmC$_A$* phase compared to SmC* in all mixtures. It means that in the SmC$_A$* phase, the average molecular shape is more deviated from the rod-like one than in the SmC* phase. However, the increase in the layer spacing deep in SmC$_A$* suggests that molecules are returning to a more extended conformation on further cooling. The $\delta\Theta$ parameter both in SmC* and SmC$_A$* shows an increasing trend with the increasing C$_m$H$_{2m}$ chain length. The recent DFT calculations for similar compounds, combined with XRD results, indicate that the CF$_2$-CF$_2$-CH$_2$-O and CF$_2$-CH$_2$-O-CH$_2$ parts of the achiral chain are in *gauche* conformation [45]. The MHPOBC molecules do not contain fluorine atoms and the XRD results for the crystal phase show that the *gauche* conformation appears only in the very end of the achiral chain, while other parts are extended [49,50]. The MHPOBC molecule has a comparable size with the 3F4HPhF6 molecule and is more likely to hinder bending of the fluorinated chain than in case of longer 3FmHPhF6 molecules with m = 5, 6 (see comparison of the achiral chain lengths in Figure 1; the size of the aromatic core and chiral chain is the same for all molecules).

The helical pitch $p_h$ in the SmC* phase of MIX4HF6 and MIX6HF6 is weakly dependent on temperature and equals ca. 325-335 nm and 420-440 nm, respectively (Figure 5d). The selectively reflected light with $\lambda \approx$ 500-650 nm is within the visible range (see Figure S13 in the Supplementary Materials). In MIX5HF6, the SmC* phase has a narrow temperature range and $p_h$ was not measured. Just below the SmC*/SmC$_A$* transition, the helical pitch equals 900-950 nm and increases with decreasing temperature for all mixtures. Thus, the selectively reflected light with $\lambda >$ 1350 nm is within the NIR range. The extrapolated temperature of the helix inversion ($p_h \to \infty$) in MIXmHF6 is 305, 338, 284 K for m = 4, 5, 6, respectively. The second branch of the $p_h(T)$ dependence, decreasing with decreasing temperature, is observed only for MIX5HF6, which has the highest inversion temperature. Below 293 K, the selectively reflected light wavenumber is in the visible range.

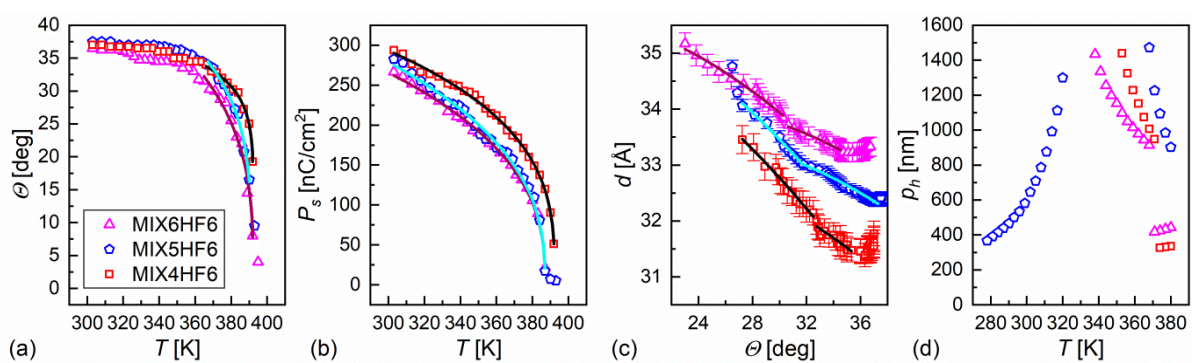

Figure 5. Tilt angle (a) and spontaneous polarization (b) vs. temperature, smectic layer spacing vs. tilt angle (c), and helix pitch vs. temperature, determined for MIXmHF6 with m = 4, 5, 6. The legend in (a) applies to all panels.

Table 2. Molecular lengths and deviations from the linear shape estimated from the plot of the smectic layer spacing vs. tilt angle for MIXmHF6.

| sample | MIX4HF6 | | MIX5HF6 | | MIX6HF6 | |
|---|---|---|---|---|---|---|
| T [K] | 347-372 | 373-387 | 315-373 | 373-382 | 339-367 | 368-383 |
| L [Å] | 33.2(3) | 36.1(3) | 33.50(4) | 36.4(2) | 34.07(5) | 35.80(9) |
| $\delta\Theta$ [deg] | 17(2) | 5(1) | 22.0(3) | 6.8(8) | 22.1(4) | 11.4(5) |

### 3.3. Dielectric relaxation processes

The investigated mixtures show the same relaxation processes, which are indicated in the representative dielectric spectra of MIX6HF6 in Figure 6. Each process is described by the relaxation time $\tau$, dielectric strength $\Delta\varepsilon$, and shape parameters $a$ and $b$. The complex function, used to fit the experimental spectra, has a form [51]:

$$\varepsilon^*(f) = \varepsilon_\infty + \sum_j \frac{\Delta\varepsilon_j}{\left(1+(2\pi i f \tau_{HNj})^{1-a_j}\right)^{b_j}} - \frac{iS}{(2\pi f)^{n_S}}, \quad (3)$$

where: the real part $\varepsilon'$ and imaginary part $\varepsilon''$ are the dielectric dispersion and absorption, respectively, $S$ and $n_S$ describe the conductivity at low frequencies, and $\varepsilon_\infty$ is the dielectric dispersion at frequency $f \to \infty$. The relaxation time $\tau_{HN}$ is equal to $\tau_{peak} = 1/2\pi f_{peak}$, where $f_{peak}$ corresponds to the peak position of the dielectric absorption, only if the absorption peak is symmetric in the $\log_{10} f$ scale. It occurs when the relaxation process is described by the Debye model ($a = 0$, $b = 1$) [51] or Cole-Cole model ($a \in (0,1)$, $b = 1$) [52]. If the relaxation process is described by the Cole-Davidson ($a = 0$, $b \in (0,1)$) [53] or the most general Havriliak-Negami model ($a \in (0,1)$, $b \in (0,1)$) [54], then the relationship between $\tau_{HN}$ and $\tau_{peak}$ is as follows [51]:

$$\tau_{peak} = \tau_{HN} \left(\sin\left(\frac{\pi(1-a)}{2+2b}\right)\right)^{-\frac{1}{1-a}} \left(\sin\left(\frac{\pi(1-a)b}{2+2b}\right)\right)^{\frac{1}{1-a}}. \quad (4)$$

The dielectric strength of each relaxation process is plotted vs. temperature in Figure 7. The dielectric dispersion at 1 kHz is shown to visualize better the phase transitions. The relaxation time of each process is plotted in the activation plot in Figure 8. Most of processes are described by the Cole-Cole model and $\tau_{peak} = \tau_{HN}$. The exception is the α-process, which is described by the Havrilak-Negami model. In this case, $\tau_{peak}$ calculated from Equation (4) is shown in the activation plots.

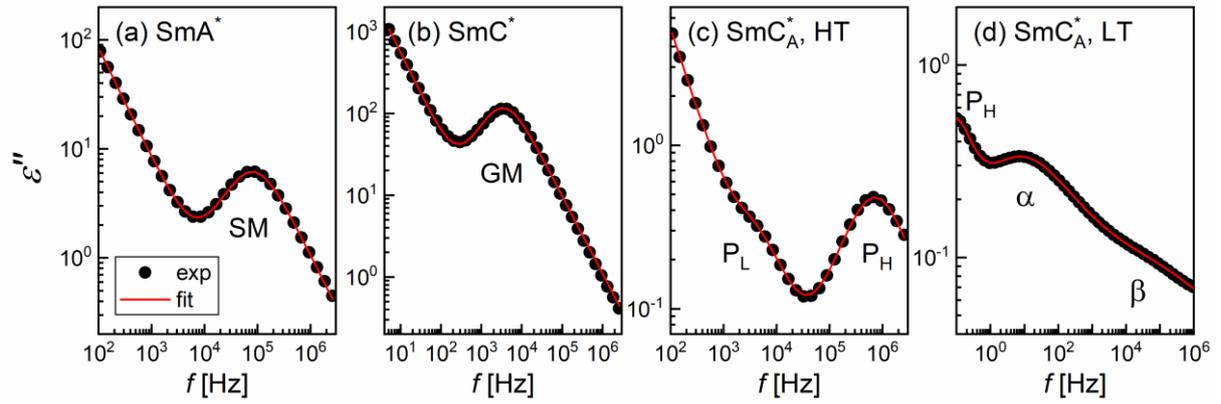

Figure 6. Dielectric absorption of MIX6HF6 vs. frequency measured on cooling and fitting results of Equation (3) in the SmA* phase in 385 K (a), SmC* phase in 381 K (b), SmC$_A$* phase in 343 K (high temperature, HT) (c), and SmC$_A$* phase in 237 K (low temperature, LT), on approaching the glass transition (d). The legend in (a) applies to the whole figure.

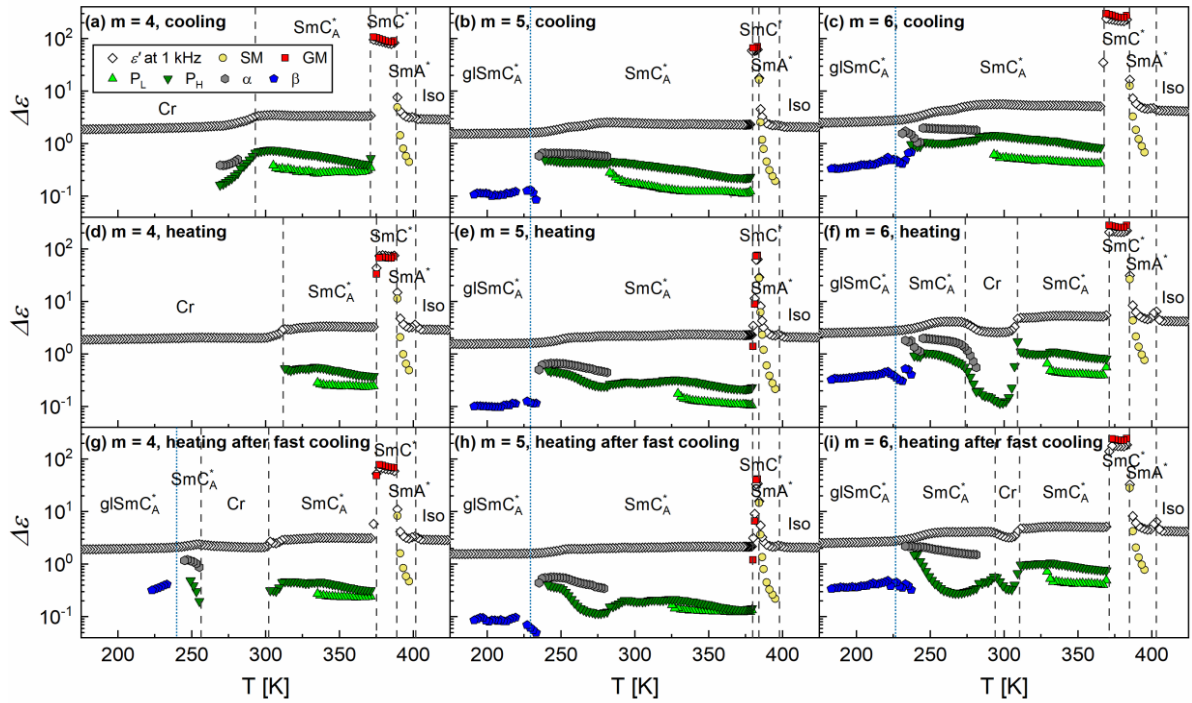

Figure 7. Dielectric strength of relaxation processes and dielectric dispersion at 1 kHz in MIXmHF6. Results in (a-c) – slow cooling, (d-f) – slow heating after slow cooling, (g-i) – slow heating after fast cooling. Panels (a,d,g) – MIX4HF6, panels (b,e,h) – MIX5HF6, panels (c,f,i) – MIX6HF6. The legend in (a) applies to the whole figure.

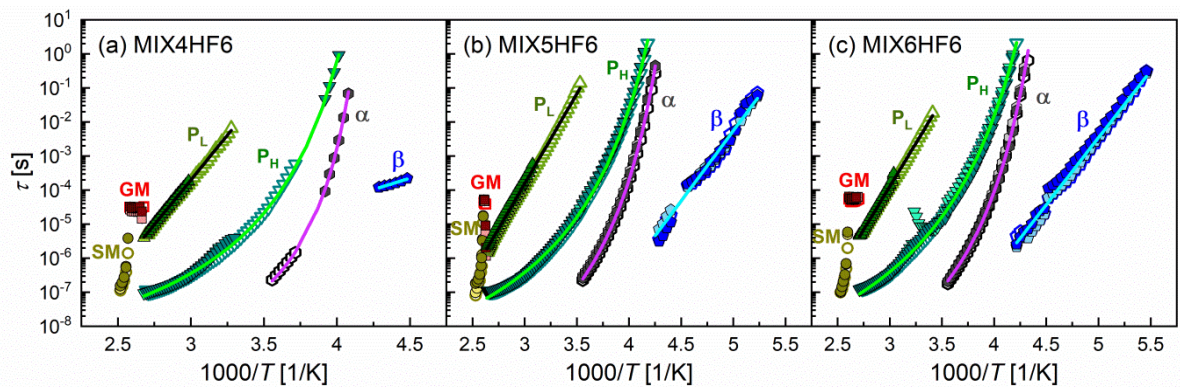

Figure 8. Activation plots of the relaxation processes in the MIXmHF6 mixtures with m = 4 (a), 5 (b), and 6 (c). Open symbols denote results for slow cooling, light solid symbols – results for heating after slow cooling, dark solid symbols – results for heating after fast cooling.

The molecular tilt in the smectic phases is described by its magnitude – the angle between the molecular long axis and the layer normal, and phase – the angle describing the orientation around the tilt cone. The relaxation processes related to the fluctuations of the magnitude and phase of the tilt are amplitudons and phasons, respectively [55]. The soft mode (SM), an amplitudon, is present in the SmA* and SmC* phase. As the temperature increases, the dielectric strength and relaxation time of SM increase in SmC* and decrease in SmA* [56,57]. The Goldstone mode (GM), a phason, is present in the SmC* phase. It is usually much stronger than SM and often makes it weakly visible in SmC*, but it can be suppressed by the external constant electric field [56,57]. The $P_L$ and $P_H$ phasons, weaker than GM, are typical for the SmC$_A$* phase. They correspond to in-phase and anti-phase fluctuations, respectively, of the tilt azimuth in neighbor smectic layers. $P_L$ has a longer relaxation time than $P_H$ [58,59]. Both $P_L$ and $P_H$ processes are initially strengthened by the external constant electric field, but at high enough voltage they are suppressed, which confirms their relationship to GM [59]. All SM, GM, $P_L$, and $P_H$ processes are observed in MIXmHF6 and confirm the identification of the smectic phases as SmA*, SmC*, and SmC$_A$* (Figure 6a-c). The inverted dielectric strength of SM is predicted to change linearly in SmA* [56]. The deviation from this dependence upon approaching the transition to SmC* can be a signature of the additional SmC$_\alpha$* phase [60]. However, the $1/\Delta\varepsilon$ vs. temperature plots of SM in MIXmHF6, shown in Figure S14 in the Supplementary Materials, are linear and do not indicate the presence of SmC$_\alpha$*.

The α-relaxation observed at higher frequencies than $P_H$ (Figure 6d) is characteristic for glassformers [61-64]. In liquid crystalline glassformers, it can be attributed to rotations around the short [65] or long [66] molecular axis (s-process or l-process). The results from [58] show that the s-process in the SmC$_A$* phase appears at lower frequencies and overlaps with the $P_L$ process, thus, the α-relaxation is more likely the l-process. The highest-frequency process is the secondary β-relaxation, originating either from movements of rigid molecules (Johari-Goldstein process) or intra-molecular motions (pseudo-JG process) [67,68]. The pseudo-JG process is more probable in MIXmHF6 due to the presence of two flexible chains in MHPOBC and 3FmHPhF6 molecules.

The $P_L$ and β-relaxation times change with temperature according to the Arrhenius formula $\tau(T) = \tau_0 \exp(E_a/RT)$ [61-64], where: $\tau_0$ is the pre-exponential constant, $E_a$ is the activation energy, and $R$ is the gas constant. The fitting results are presented in Table 3. The $P_L$ process shows quantitatively similar dependence on temperature for all mixtures, with $\tau_0 \approx 10^{-19}$ s and $E_a \approx 100$ kJ/mol. The β-relaxation is characterized by $\tau_0 = 10^{-23}$-$10^{-22}$ s and $E_a = 75$-$80$ kJ/mol for MIX5HF6 and MIX6HF6. The results are very different for MIX4HF6, where $\tau_0 \approx 10^{-10}$ s and $E_a = 26$ kJ/mol. MIX4HF6 crystallizes on slow cooling and the β-relaxation was investigated only upon heating after fast cooling (Figure 8a), in a temperature range much narrower than for m = 5, 6 (Figure 8b,c), which may contribute to the observed difference. The $P_H$ and α-relaxation time changes according to the Vogel-Fulcher-Tammann formula $\tau(T) = \tau_0 \exp(B/(T-T_V))$ [61-64], where $T_V$ is the Vogel temperature and $B$ is a parameter which corresponds to $E_a/R$ in the limit of $T_V \to 0$. The fitting results are presented in Table 4. The obtained parameters of the VFT equation are comparable for m = 5, 6, while for m = 4, the $B$ value is lower and $T_V$ is higher both for the $P_H$ and α-relaxations. The deviation of the α-relaxation time from the Arrhenius formula is expressed by the fragility parameter $m_f$, which is defined as [61]:

$$m_f = \frac{d \log_{10} \tau_\alpha}{d(T_g/T)}\bigg|_{T=T_g}, \qquad (5)$$

where: the glass transition temperature $T_g$ is defined as the temperature where the α-relaxation time equals 100 s [61]. The $T_g$ temperature determined from the VFT parameters for the α-relaxation increases with increasing $C_mH_{2m}$ chain length and is in agreement with the DSC results (Table 1). The fragility parameter is $m_f = 111$-$112$ for m = 5, 6 and 188 for m = 4. Higher fragility of the glassformer is sometimes interpreted as a lower glassforming ability [69], which agrees with the results for MIXmHF6. Although $T_g$ and $m_f$ should be investigated based on the α-relaxation time, the corresponding values were obtained also for $P_H$ for comparison, as it was done in [70]. The glass transition temperature determined from the $P_H$ relaxation time is only 3-4 K higher than that obtained from the α-relaxation time and can be a good estimation when the α-relaxation cannot be observed. On the contrary, $m_f$ values obtained from the $P_H$ and α-relaxations differ considerably, thus, $P_H$ relaxation time should not be used for estimation of fragility.

Table 3. Fitted parameters of Arrhenius formula for the $P_L$ and β-relaxation times in MIXmHF6.

| mixture | process | $\log_{10}(\tau_0/s)$ | $E_a$ [kJ/mol] |
|---|---|---|---|
| MIX4HF6 | $P_L$ | −19.5(2) | 100.5(8) |
|  | β | −9.7(5) | 26(2) |
| MIX5HF6 | $P_L$ | −19.6(2) | 101.0(8) |
|  | β | −22.8(4) | 79(2) |
| MIX6HF6 | $P_L$ | −19.08(6) | 96.9(4) |
|  | β | −22.1(3) | 75.2(9) |

Table 4. Fitted parameters of VFT formula and derived glass transition temperatures $T_g$ and fragility indices $m_f$ of MIXmHF6 based on the $P_H$ and α-relaxation times.

| mixture | process | $\log_{10}(\tau_0/s)$ | $B$ [K] | $T_V$ [K] | $T_g$ [K] | $m_f$ |
|---|---|---|---|---|---|---|
| MIX4HF6 | $P_H$ | −9.09(3) | 722(9) | 214.5(3) | 242.8(1) | 95.2(8) |
|  | α | −9.8(1) | 410(11) | 224.7(2) | 239.8(1) | 188(2) |
| MIX5HF6 | $P_H$ | −9.41(4) | 933(12) | 197.8(5) | 233.3(1) | 74.9(6) |
|  | α | −11.0(2) | 798(38) | 202.7(8) | 229.3(2) | 112(2) |
| MIX6HF6 | $P_H$ | −9.60(3) | 1045(10) | 191.4(3) | 230.6(1) | 68.4(4) |
|  | α | −10.8(1) | 768(17) | 200.6(4) | 226.6(1) | 111.4(9) |

The relaxation time $\tau_G$ and dielectric strength $\Delta\varepsilon_G$ of the Goldstone mode can be combined with the tilt angle $\Theta$ and spontaneous polarization $P_s$ to obtain the rotational viscosity in SmC* [71]:

$$\gamma_{rot} = \frac{P_s^2 \tau_G}{2\varepsilon_0 \Theta^2 \Delta\varepsilon_G}, \quad (6)$$

where $\varepsilon_0$ is the vacuum permittivity. The rotational viscosity in the SmC* phase of MIX4HF6 and MIX6HF6 changes according to the Arrhenius formula (Figure 9) and the corresponding activation energies are equal to 29(3) kJ/mol and 28(3) kJ/mol, respectively. The activation plot of $\gamma_{rot}$ deviates from the Arrhenius formula for MIX5HF6, probably due to a narrow temperature range of SmC*, and the activation energy was not determined.

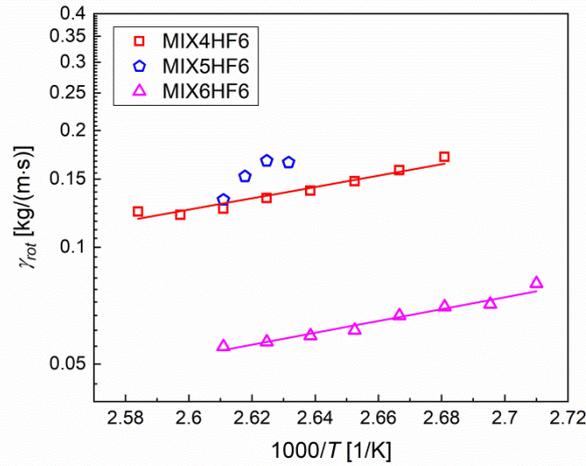

Figure 9. Activation plot of the rotational viscosity in the SmC* phase of MIXmHF6.

## 4. Summary and conclusions

The partially fluorinated 3FmHPhF6 homologs with m = 4, 5, and 6 were mixed with MHPOBC to formulate the equimolar mixtures with the corresponding acronyms MIXmHF6. The complementary DSC, POM, XRD, and BDS methods indicate the phase sequence SmA* → SmC* → SmC$_A$* on cooling for all mixtures. The length of the C$_m$H$_{2m}$ achiral chain in the 3FmHPhF6 component influences properties of the mixtures in various ways, namely:

- the smectic layer spacing increases, the layer shrinkage, spontaneous polarization, and glass transition temperature decrease, and the estimated shape of the molecules deviates stronger from the rod-like one with the increasing m parameter,
- the critical cooling rate necessary to obtain the SmC$_A$* glass is lower, the inversion temperature of the helix in the SmC$_A$* phase is higher, and the range of the SmC* phase is narrower for m = 5 than for m = 4 and 6 (odd-even effect),
- the fragility index of the supercooled SmC$_A$* and the melting temperature are comparable for m = 5, 6 and higher for m = 4,
- the tilt angle does not depend significantly on the m parameter.

From the practical point of view, the best properties are exhibited by MIX5HF6, which does not crystallize on slow cooling and has the lowest tendency to the cold crystallization on heating. Moreover, only in MIX5HF6 the thermochromic property – selective reflection of light with a wavenumber strongly dependent on temperature – is observed both in the visible range and close to the room temperature.

**Acknowledgement:** We thank Dr. Eng. Magdalena Urbańska from the Institute of Chemistry of the Military University of Technology in Warsaw for chemical synthesis. Aleksandra Deptuch acknowledges the National Science Centre, Poland (grant MINIATURA 7 no. 2023/07/X/ST3/00182) for financial support.

**Conflicts of interest statement:** There are no conflicts to declare.

**Authors' contributions:**

A. Deptuch – conceptualization, investigation, formal analysis, funding acquisition, writing – original draft

A. Paliga – investigation, formal analysis, writing – review and editing

A. Drzewicz – investigation, writing – review and editing

M. Czerwiński – investigation, writing – review and editing

E. Juszyńska-Gałązka – investigation, writing – review and editing

# Glassforming liquid crystalline equimolar mixtures of MHPOBC and fluorinated compounds – structural, optical, and dielectric properties

Aleksandra Deptuch[1,*], Anna Paliga[2], Anna Drzewicz[1], Michał Czerwiński[3], Ewa Juszyńska-Gałązka[1,4]

[1] Institute of Nuclear Physics Polish Academy of Sciences, Radzikowskiego 152, PL-31342 Kraków, Poland
[2] Faculty of Physics and Applied Computer Science, AGH University of Kraków, Reymonta 19, PL-30059 Kraków, Poland
[3] Institute of Chemistry, Military University of Technology, Kaliskiego 2, PL-00908 Warsaw, Poland
[4] Research Center for Thermal and Entropic Science, Graduate School of Science, Osaka University, 560-0043 Osaka, Japan
*corresponding author, aleksandra.deptuch@ifj.edu.pl

# Supplementary Materials

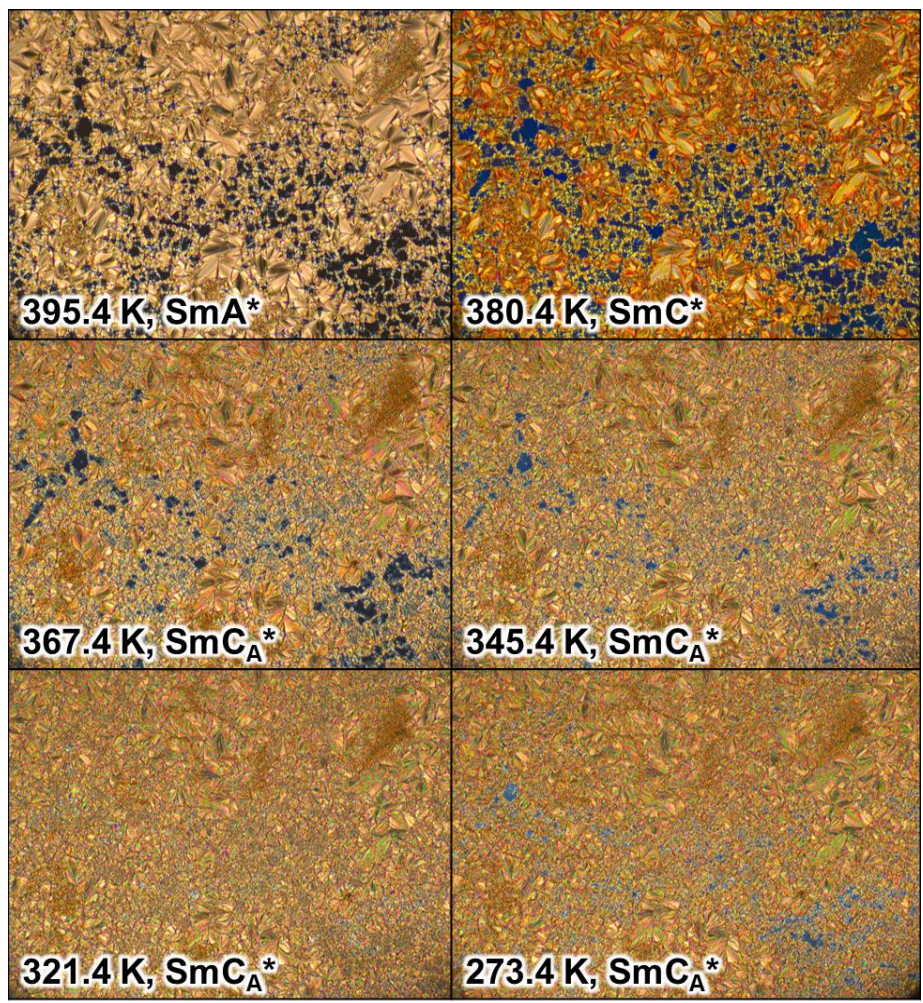

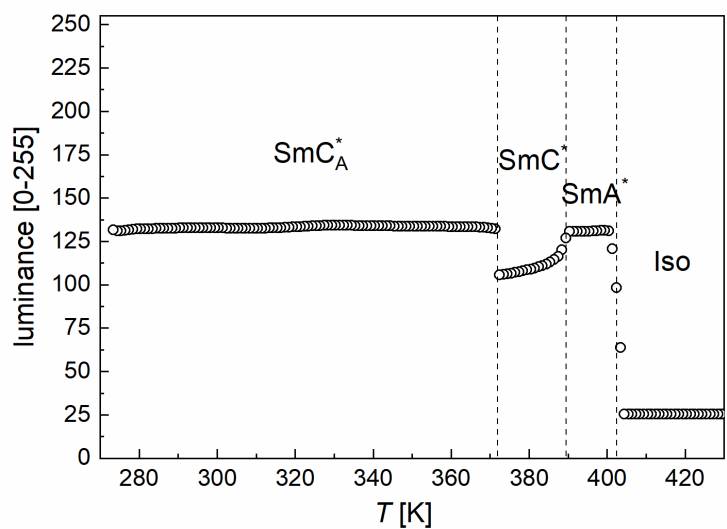

Figure S1. Selected POM textures of MIX4HF6 and average luminance of all textures collected during cooling at 5 K/min. Each texture covers an area of 1243 × 933 μm$^2$.

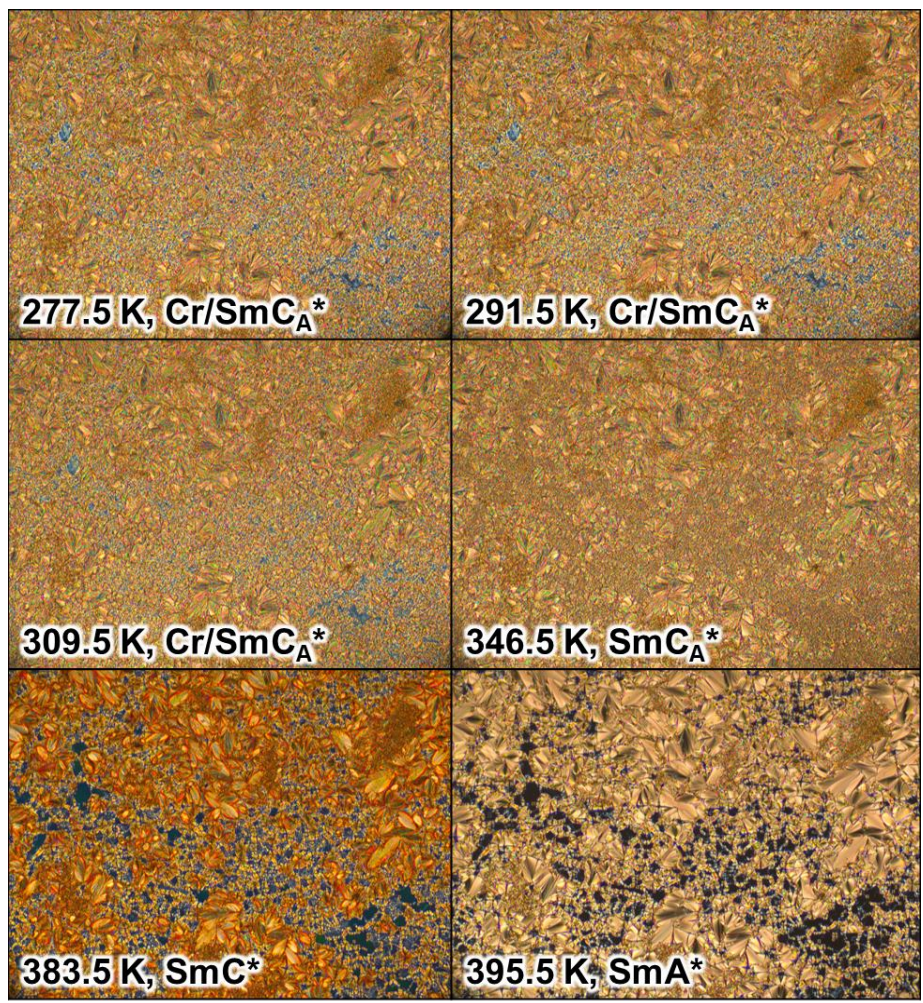

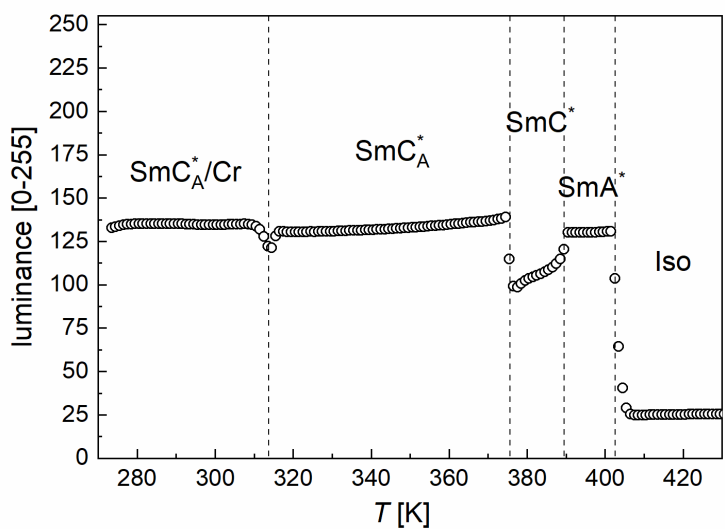

Figure S2. Selected POM textures of MIX4HF6 and average luminance of all textures collected during heating at 5 K/min. Each texture covers an area of 1243 × 933 μm$^2$.

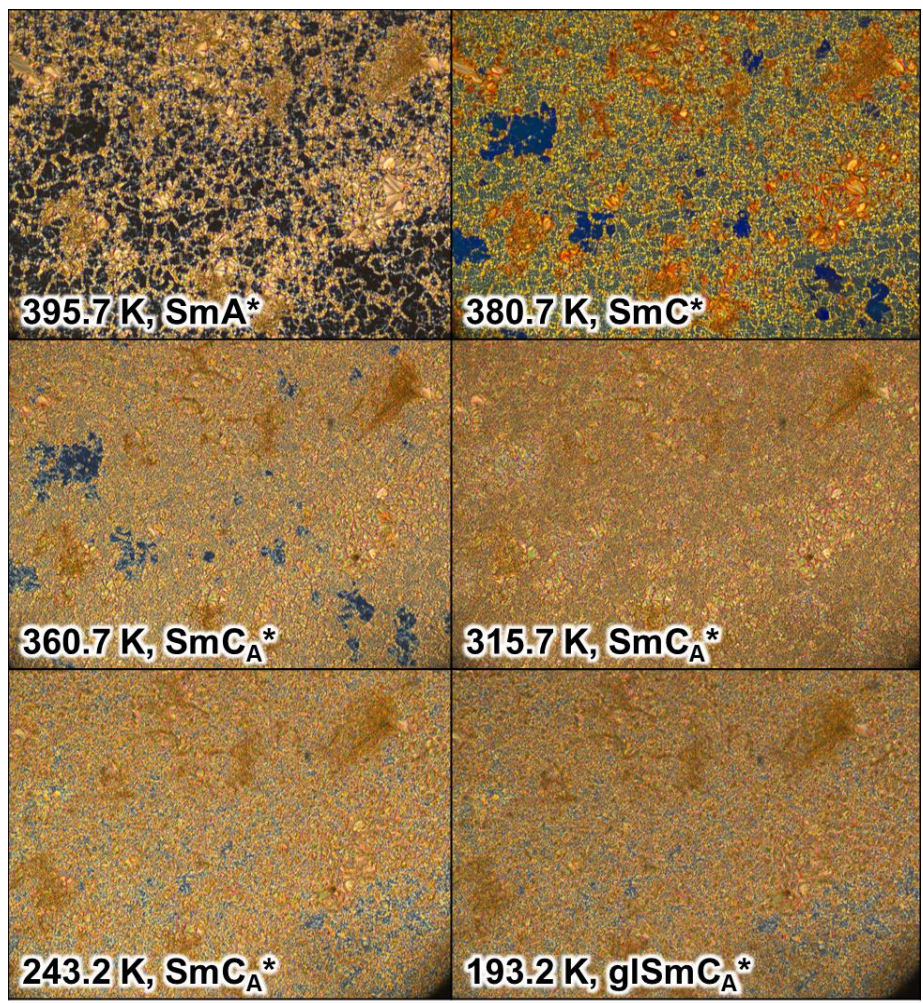

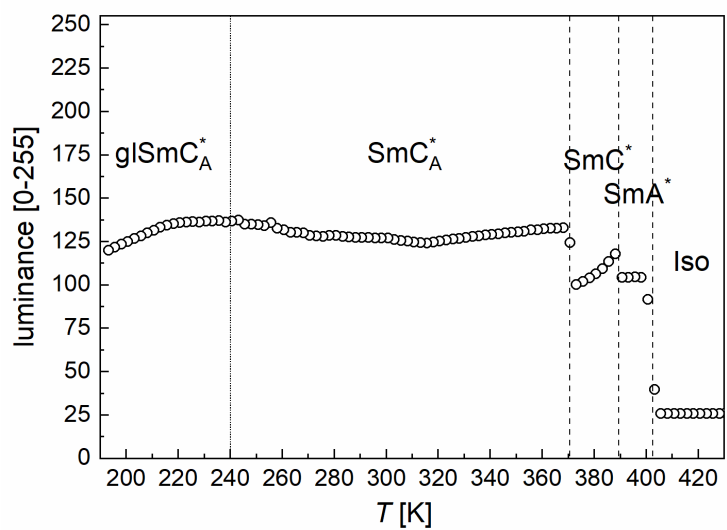

Figure S3. Selected POM textures of MIX4HF6 and average luminance of all textures collected during cooling at 30 K/min. Each texture covers an area of 1243 × 933 μm$^2$. The glass transition temperature is based on DSC results.

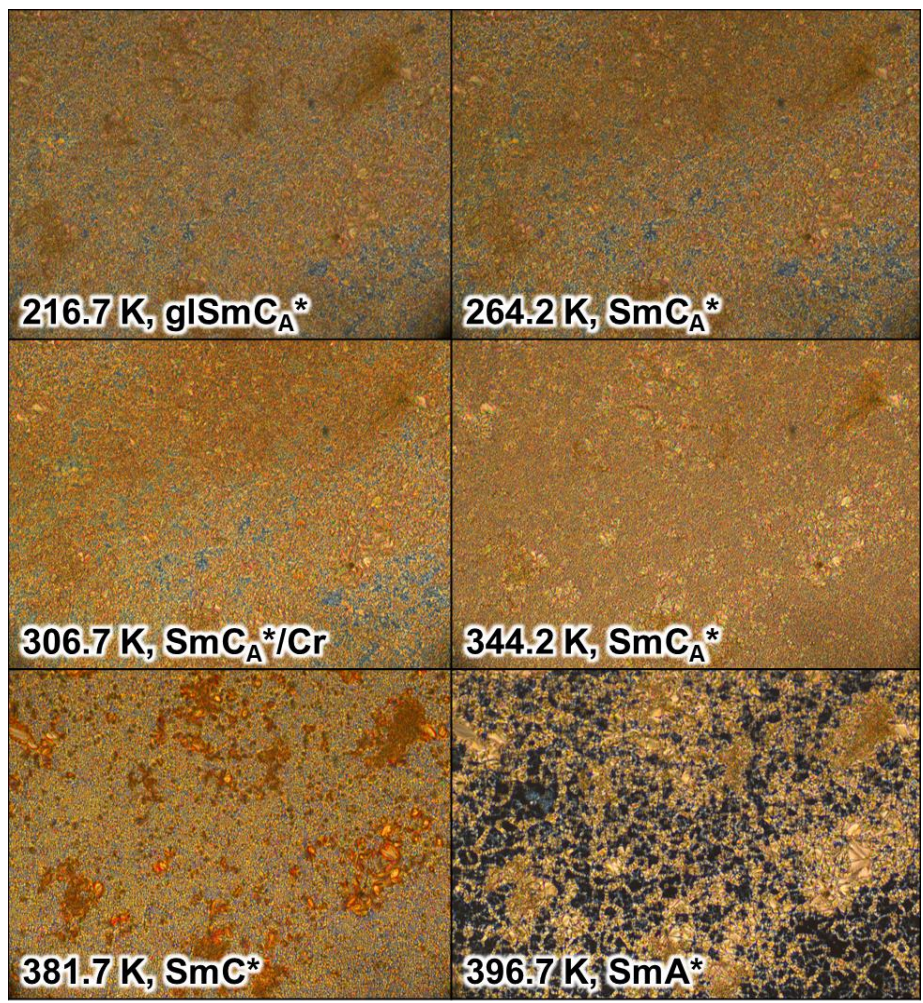

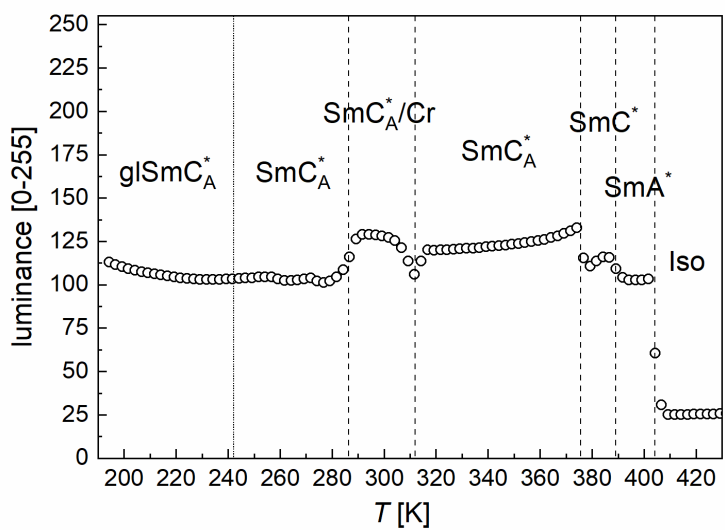

Figure S4. Selected POM textures of MIX4HF6 and average luminance of all textures collected during heating at 30 K/min. Each texture covers an area of 1243 × 933 μm$^2$. The glass transition temperature is based on DSC results.

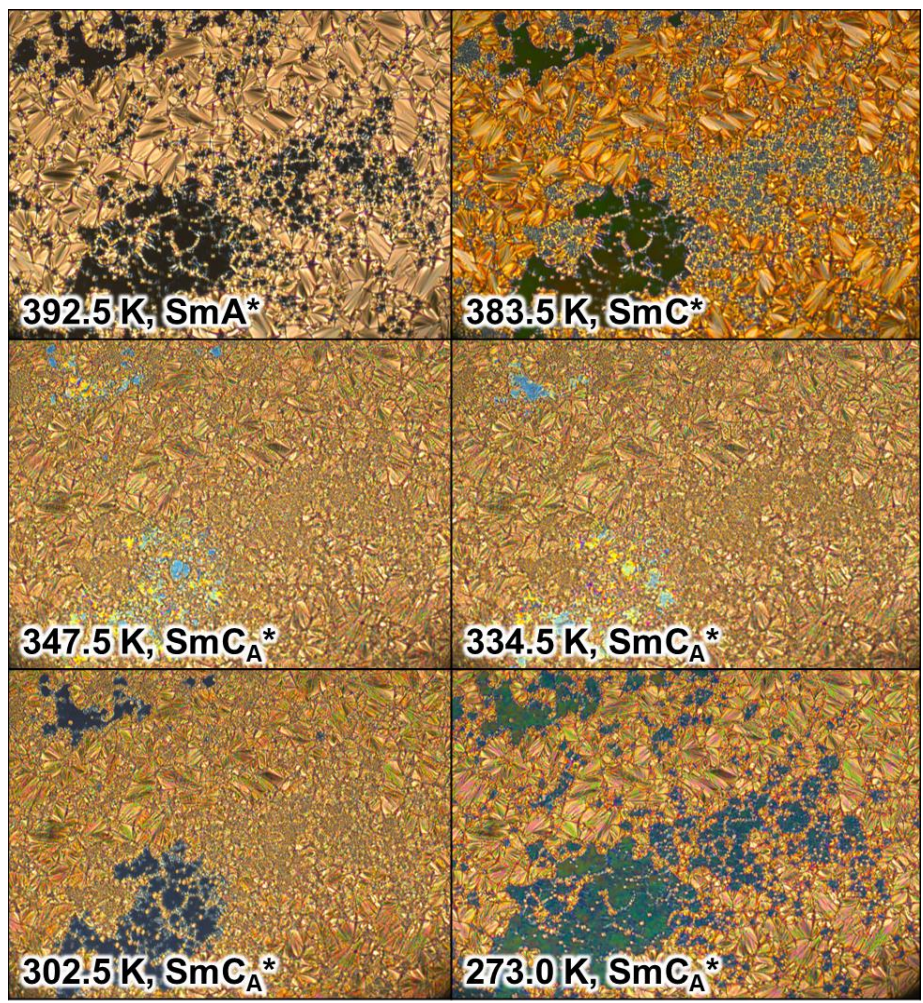

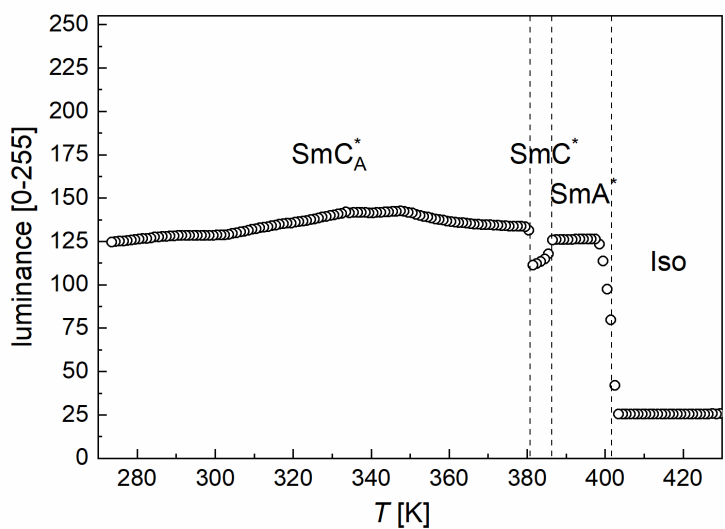

Figure S5. Selected POM textures of MIX5HF6 and average luminance of all textures collected during cooling at 5 K/min. Each texture covers an area of 1243 × 933 μm$^2$.

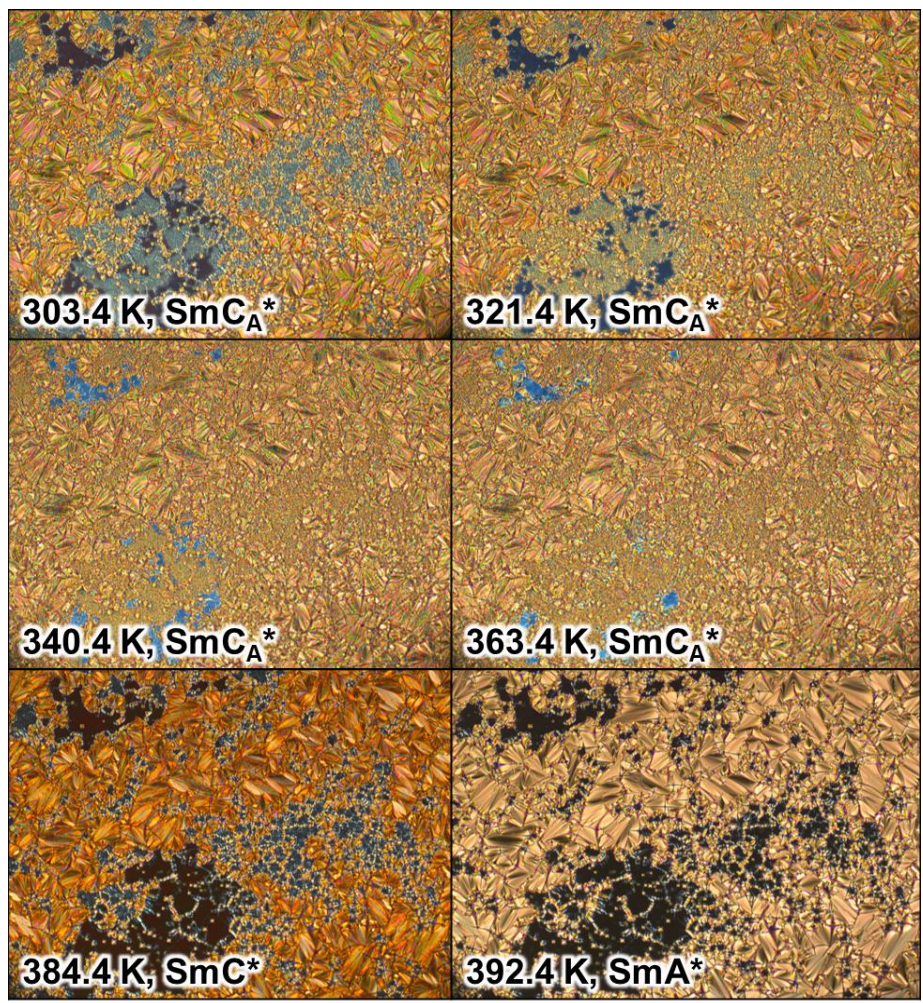

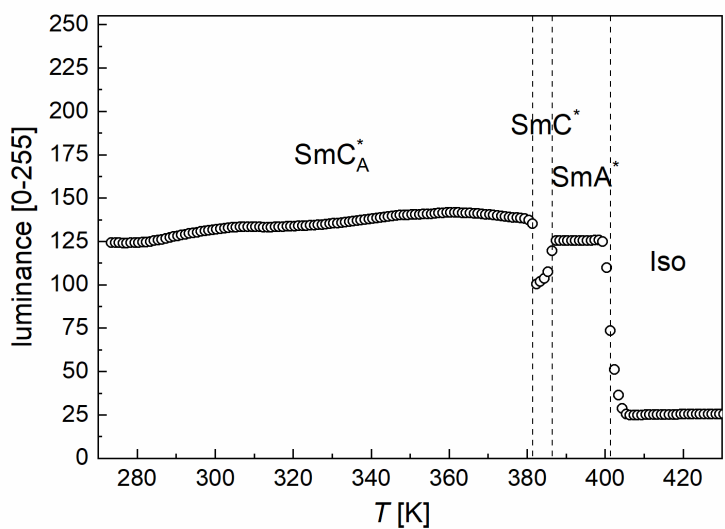

Figure S6. Selected POM textures of MIX5HF6 and average luminance of all textures collected during heating at 5 K/min. Each texture covers an area of 1243 × 933 μm$^2$.

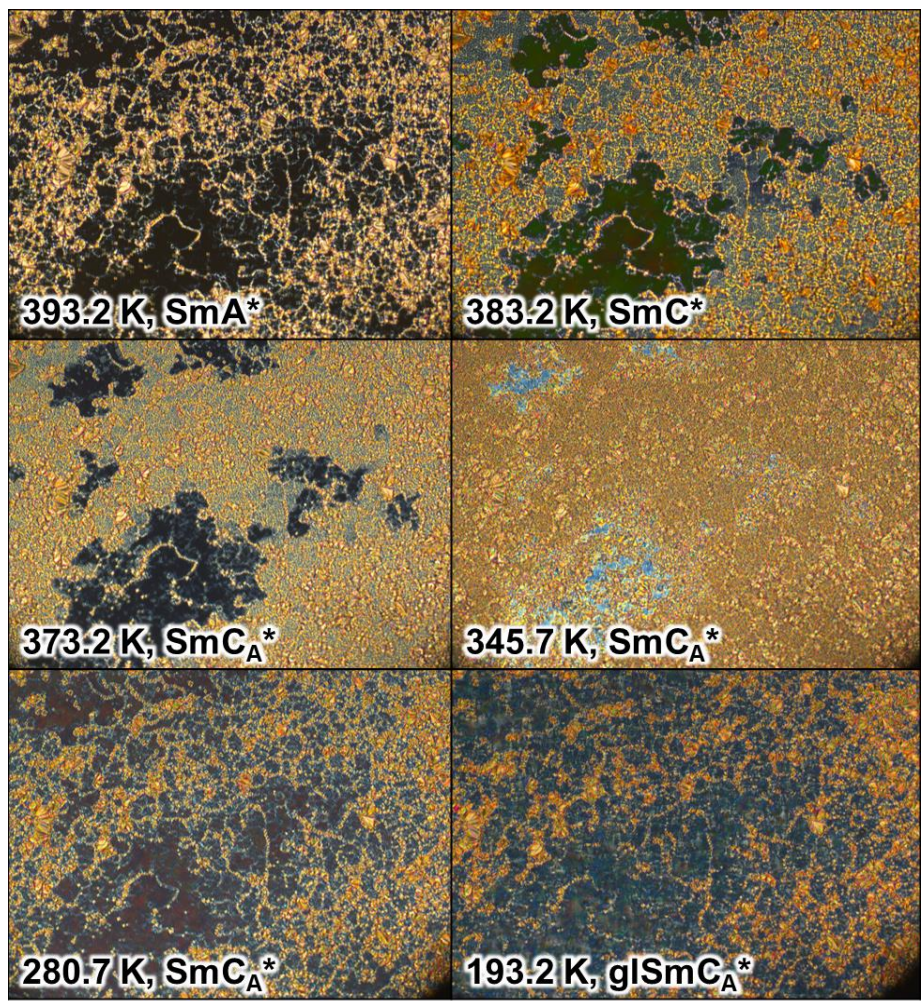

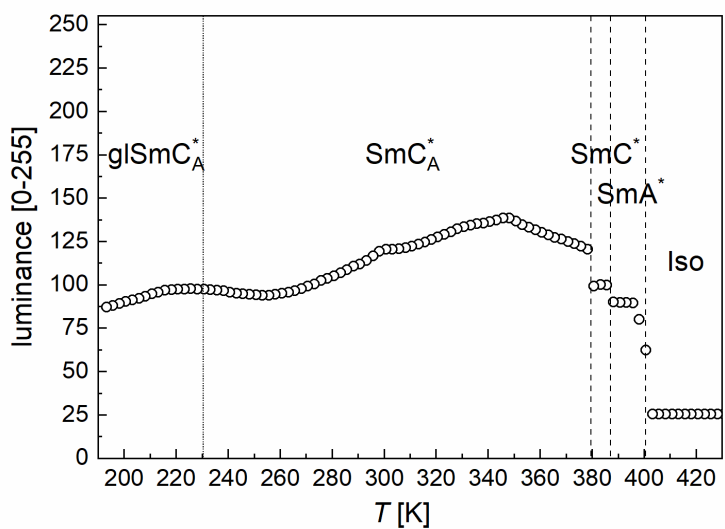

Figure S7. Selected POM textures of MIX5HF6 and average luminance of all textures collected during cooling at 30 K/min. Each texture covers an area of 1243 × 933 μm$^2$. The glass transition temperature is based on DSC results.

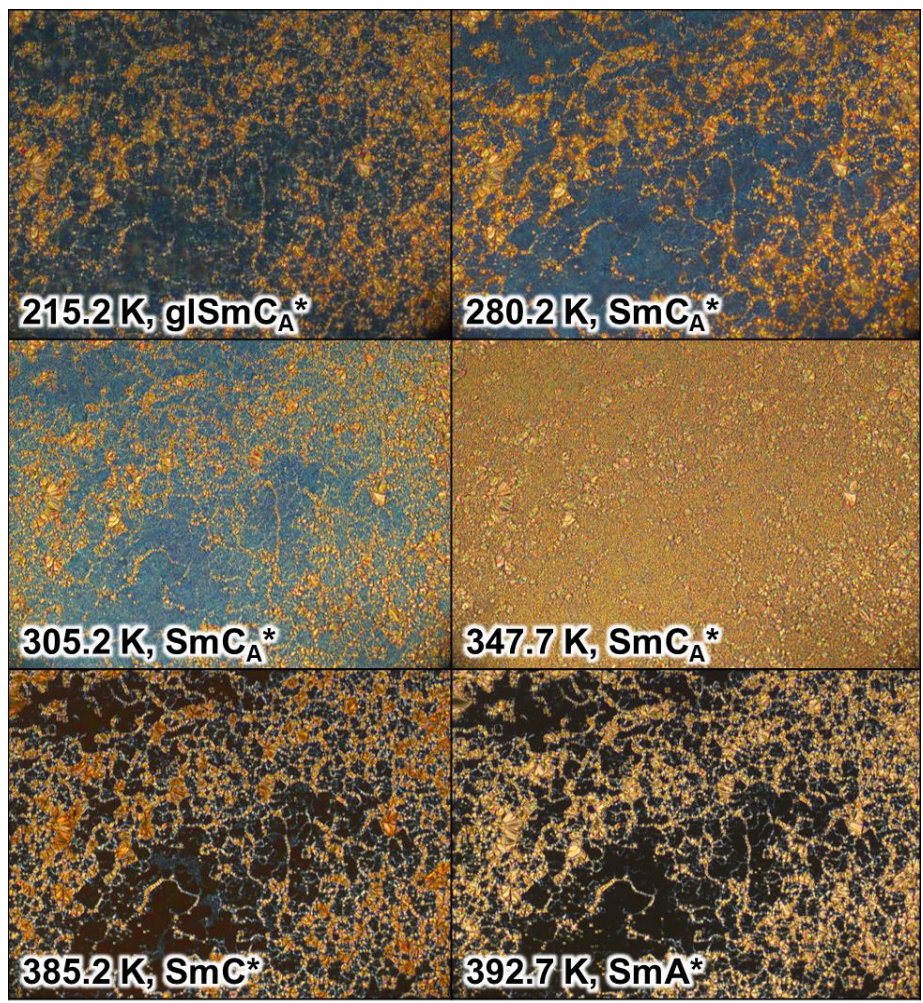
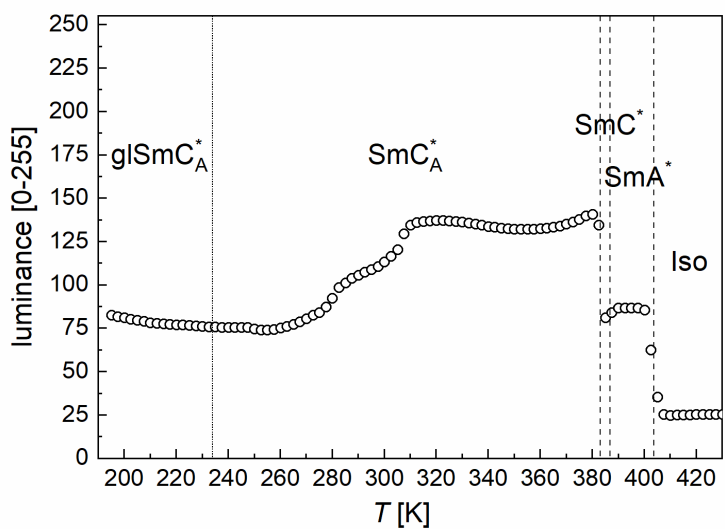

Figure S8. Selected POM textures of MIX5HF6 and average luminance of all textures collected during heating at 30 K/min. Each texture covers an area of 1243 × 933 μm$^2$. The glass transition temperature is based on DSC results.

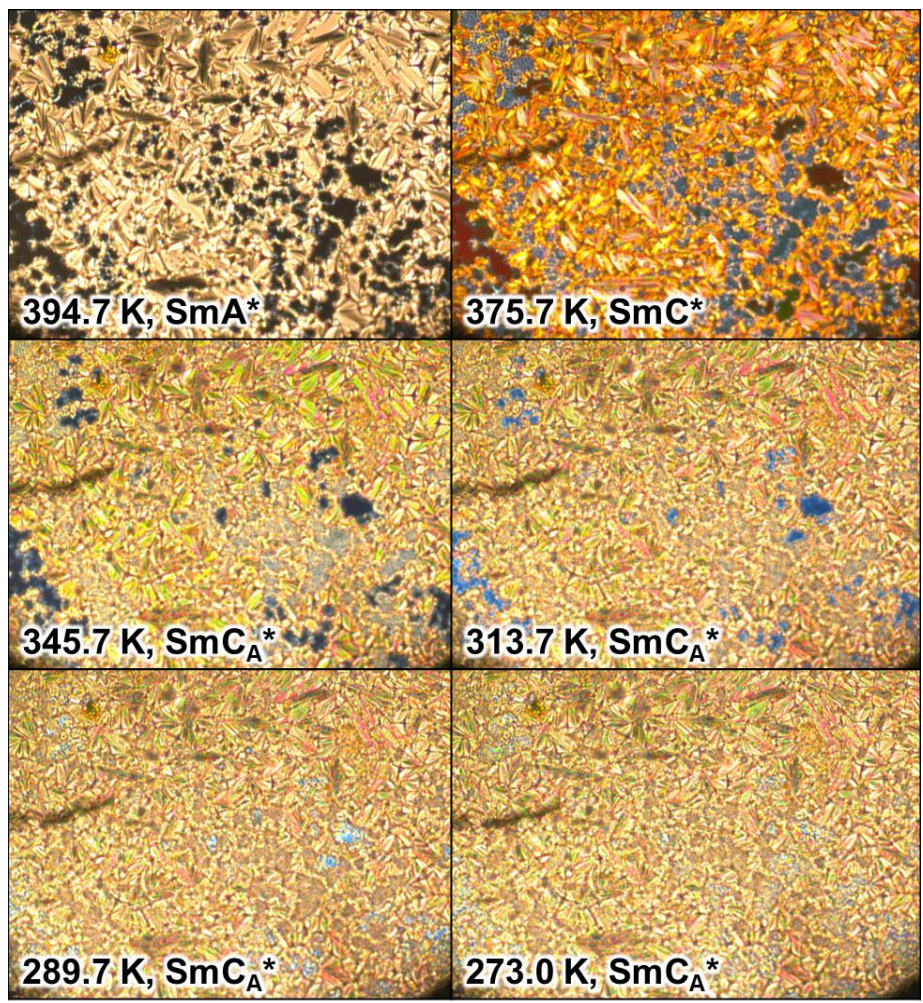

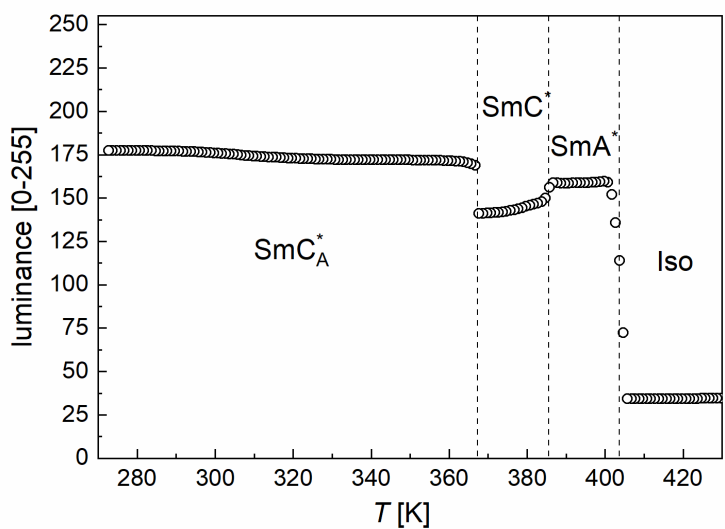

Figure S9. Selected POM textures of MIX6HF6 and average luminance of all textures collected during cooling at 5 K/min. Each texture covers an area of 1243 × 933 μm$^2$.

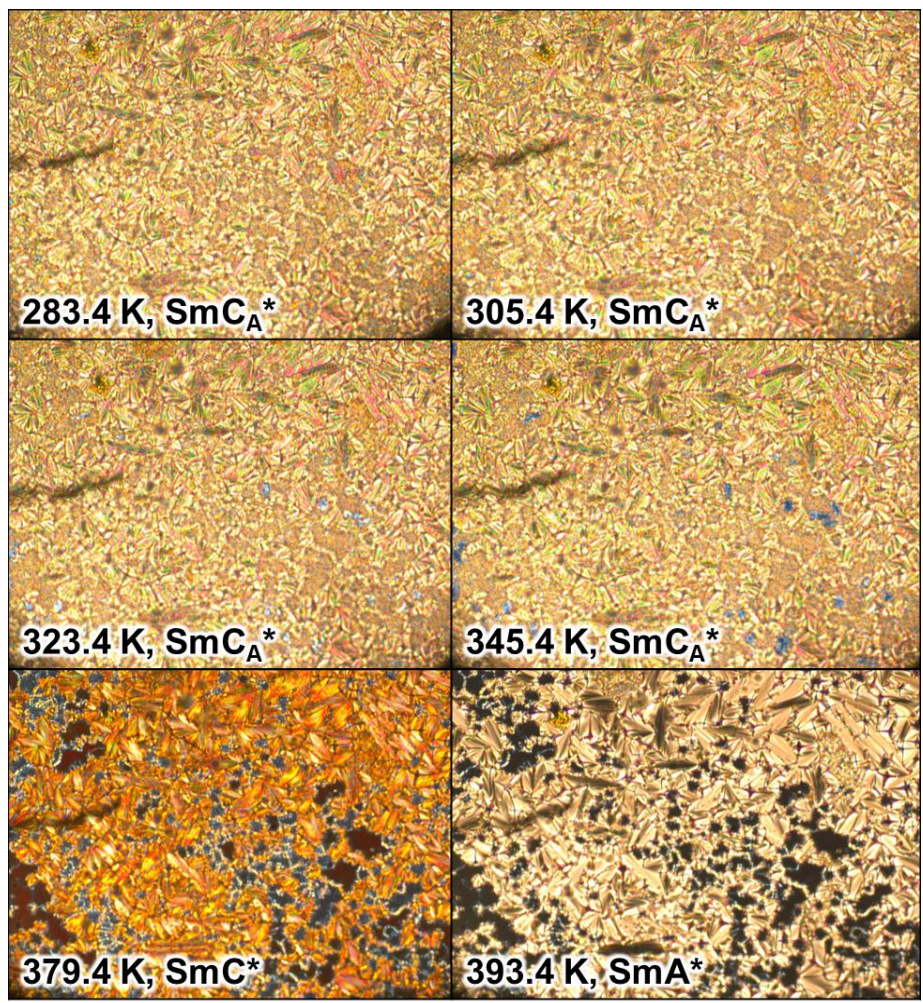

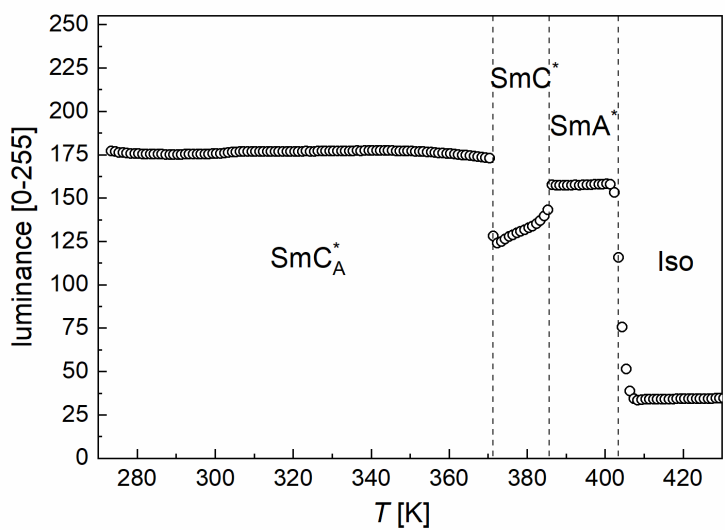

Figure S10. Selected POM textures of MIX6HF6 and average luminance of all textures collected during heating at 5 K/min. Each texture covers an area of 1243 × 933 μm$^2$.

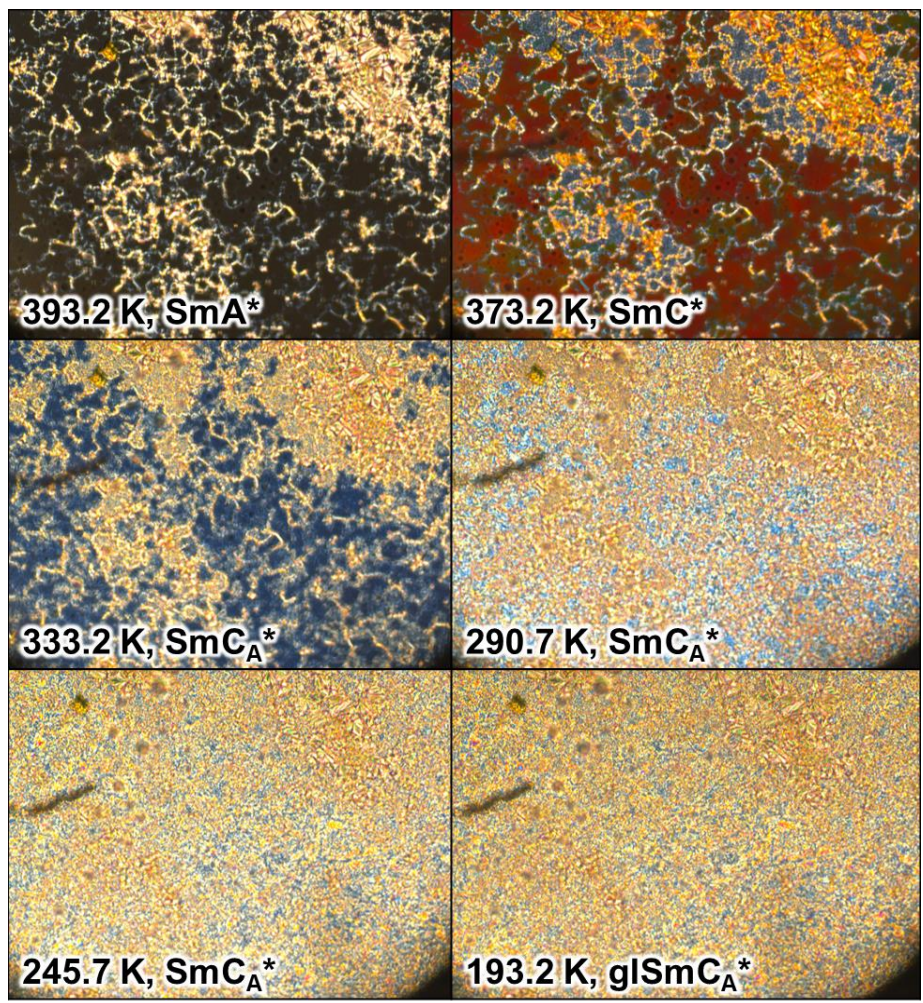

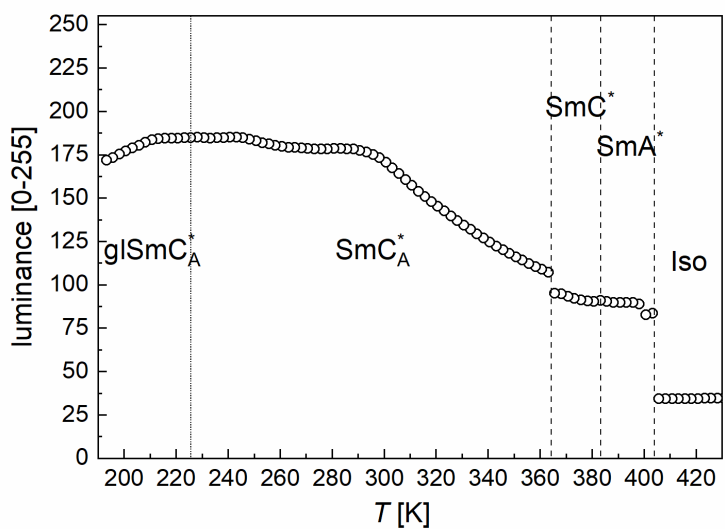

Figure S11. Selected POM textures of MIX6HF6 and average luminance of all textures collected during cooling at 30 K/min. Each texture covers an area of 1243 × 933 μm$^2$. The glass transition temperature is based on DSC results.

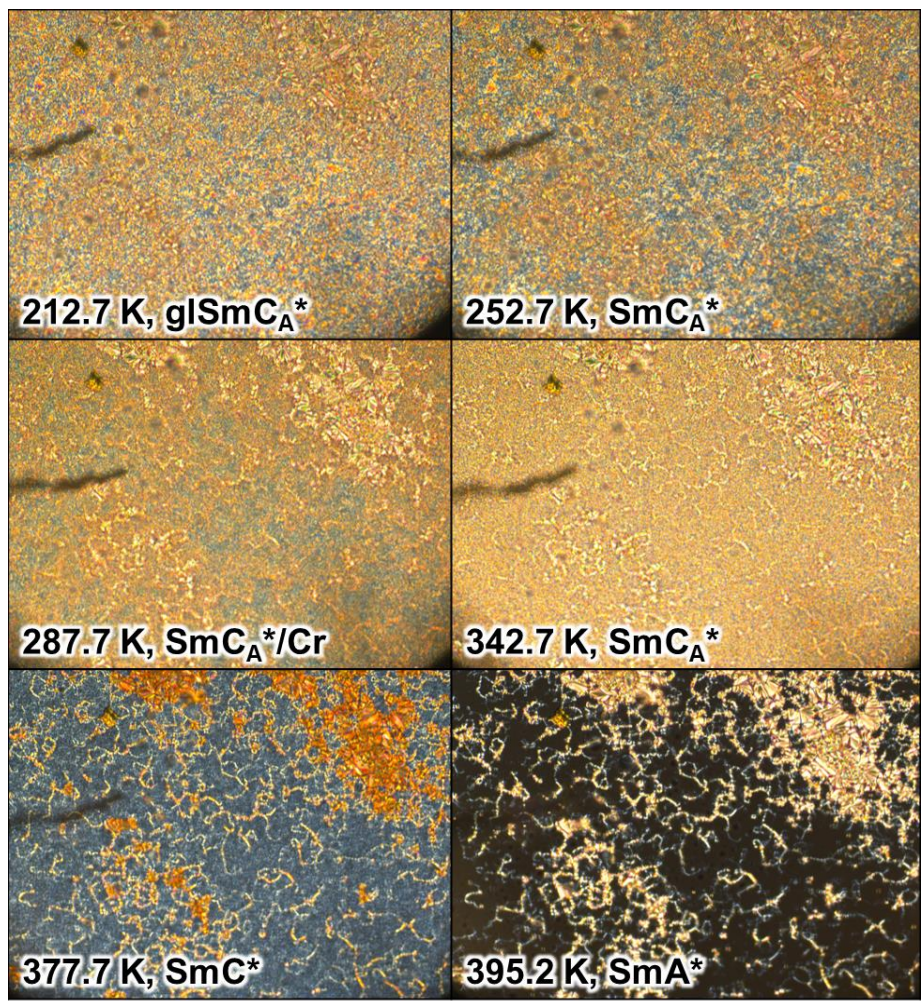

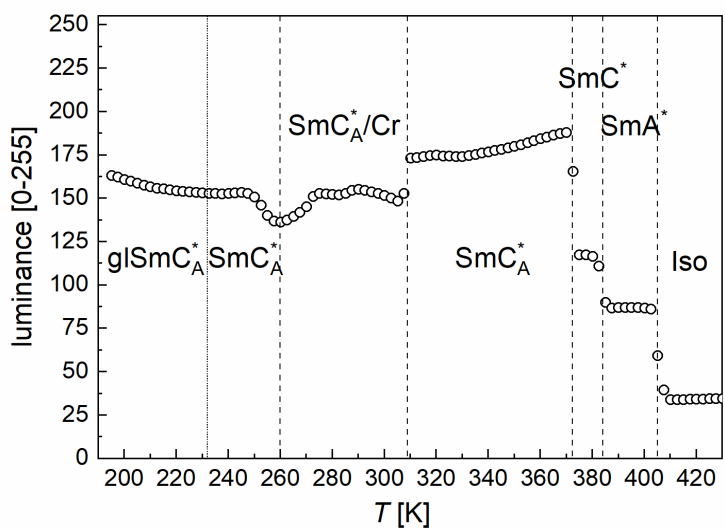

Figure S12. Selected POM textures of MIX6HF6 and average luminance of all textures collected during heating at 30 K/min. Each texture covers an area of 1243 × 933 μm$^2$. The glass transition temperature is based on DSC results.

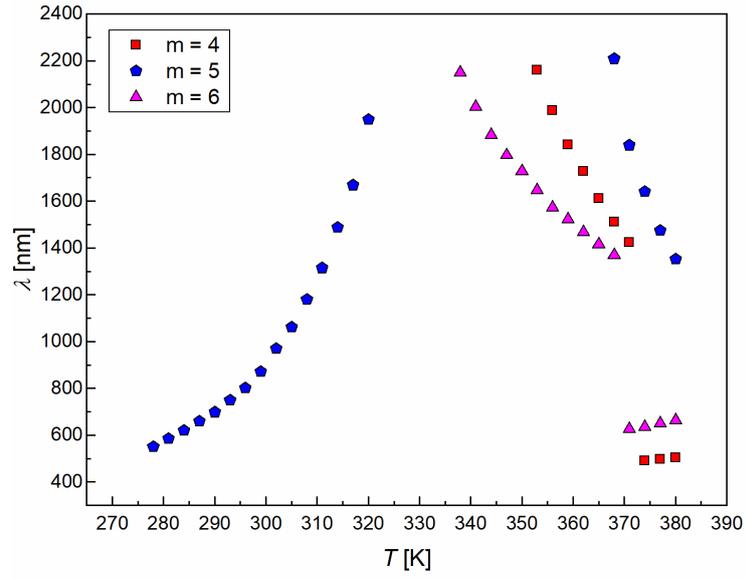

Figure S13. Wavelength of the selectively reflected light in the MIXmHF6 mixtures vs. temperature.

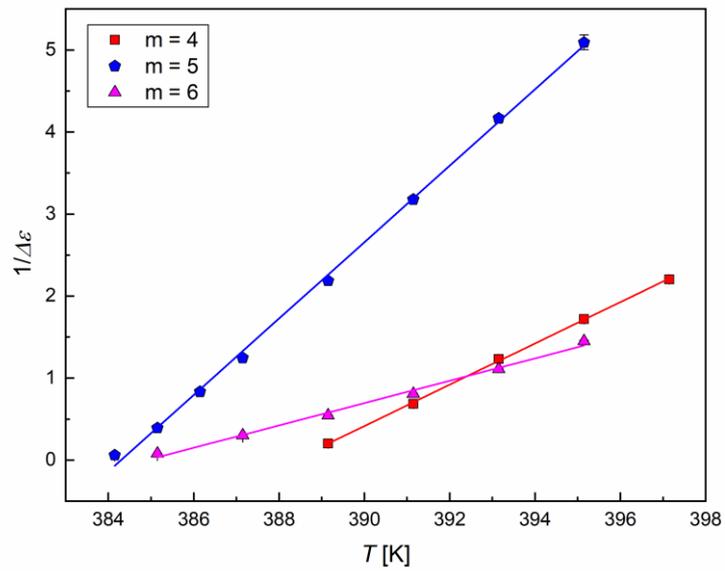

Figure S14. Inverted dielectric strength of the soft mode in the MIXmHF6 mixtures vs. temperature. The results were obtained on slow cooling.